\documentclass[%
 preprint,
 showpacs,
 preprintnumbers,
 amsmath,amssymb,
 longbibliography,
 lengthcheck,
]{revtex4-1}

\usepackage{natbib}
\usepackage[dvipdfm]{graphicx,color}
\usepackage{dcolumn}
\usepackage{bm}
\usepackage{hyperref}

\begin{document}

\title{Spin Excitation Assisted by Non-Softening Phonon \\for Spin-Peierls Model}

\author{Takanori Sugimoto}
\email{takanori@yukawa.kyoto-u.ac.jp}
\author{Shigetoshi Sota}
\author{Takami Tohyama}
\affiliation{Yukawa Institute for Theoretical Physics, Kyoto University, 606-8502, Kyoto, Japan}
\date{\today}

\begin{abstract}
We study spin dynamics of a spin-Peierls chain with nearest-neighbor and next-nearest-neighbor Heisenberg spin exchange interactions together with a gapped and dispersionless phonon. The dynamical spin correlation function and phonon excitation spectrum are calculated at zero temperature by using dynamical density-matrix renormalization-group method.  We find a new spin excitation assisted by non-softening phonon. The excitation is located above phonon in energy and shows a dispersive feature with strong intensity near the momentum $\pi$. The phonon excitation spectrum is also influenced by the spin-phonon interaction. We discuss the possibility of observing the spin-phonon coupled features in inorganic spin-Peierls compound CuGeO$_3$.
\end{abstract}

\maketitle

\section{Introduction}
One-dimensional (1D) quantum spin system coupled with lattice degree of freedom has been extensively studied experimentally and theoretically, since the systems provide a playground of spin-Peierls transition. Conventional spin-Peierls compounds like organic materials exhibit the transition with spin dimerization and lattice alternation accompanied by soft-phonon mode.~\cite{bray75,huizinga79}  Theoretically the spin-Peierls transition can be derived under the presence of soft-mode phonon by using the random phase approximation approach in the adiabatic limit.~\cite{cross79} The discovery of CuGeO$_3$,~\cite{hase93} however, has casted a problem on the conventional mechanism of the spin-Peierls transition, since the soft-phonon mode associated to lattice alternation has never been found in CuGeO$_3$ so far.~\cite{hase93,nishi94,castilla95,riera95}

A recent study has shown a theoretical explanation of spin-Peierls instability in the antiadiabatic limit. ~\cite{uhrig98} The spin-Peierls Hamiltonian can be mapped approximately to spin-$\frac{1}{2}$ $J_1$-$J_2$ Heisenberg Hamiltonian by the flow equation in the antiadiabatic limit, where $J_1$ and $J_2$ is nearest-neighbor (NN) and next-nearest-neighbor (NNN) exchange interactions, respectively.  The spin-Peierls transition, therefore, corresponds to the Kosterlitz-Thouless (KT) transition in the $J_1$-$J_2$ Heisenberg model obtained by changing the ratio of $J_1$ and $J_2$.  The KT transition is order-disorder type quantum phase transition, and has a critical exchange ratio, $\alpha_c\cong 0.241167$, between spin-liquid and dimer phases at the ground state.~\cite{affleck89,eggert96,okamoto92}  Namely, for the exchange ratio $J_2/J_1\equiv\alpha<\alpha_c$, the ground state belongs to the spin-liquid phase, but for $\alpha>\alpha_c$, it is in the dimer phase.  The renormalization of phonon degree of freedom by the flow equation changes the ratio $\alpha$: $\alpha$ increases as spin-phonon coupling $g$ increases.  Thus, there occurs the KT transition without soft phonons in the antiadiabatic limit for large spin-phonon coupling  $g>g_c\cong0.8$.~\cite{gros98,bursill99,sandvik99,barford05,weisse06,dobry07,pearson10}  This means that no soft phonons exist below the critical temperature.~\cite{uhrig98}

The spin-Peierls transition was experimentally observed for the organic compounds, tetrathiafulvalene (TTF), tetracyanoquinodimethane (TCNQ) series in the 1970s,~\cite{bray75,kasper79,huizinga79,bodegom81,visser83,aeppli84} and the inorganic compounds CuGeO$_3$ in 1993.~\cite{hase93}  Experimental studies have shown that phonon frequency associated with lattice distortion in CuGeO$_3$ is higher than that in the organic compounds. In (TTF)CuS$_4$C$_4$(CF$_3$)$_4$, a soft-phonon frequency is estimated to be $\omega_\mathrm{ph}\cong 1.4 \, \mathrm{meV}\cong 16 \,\mathrm{K}$, and spin-Peierls gap is $\Delta\cong 21 \,\mathrm{K}$.~\cite{bray75,kasper79}  On the other hand, for CuGeO$_3$, a dispersive phonon frequency related to lattice distortion is $\omega_\mathrm{ph}\cong 6.8 \,\mathrm{THz}\cong 330 \,\mathrm{K}$, a dispersionless phonon is $\omega_\mathrm{ph}\cong 3.2 \,\mathrm{THz}\cong 150 \,\mathrm{K}$, and the gap is $\Delta\cong2 \,\mathrm{meV}\cong23\,\mathrm{K}$.~\cite{nishi94,braden98-1,braden98-2,braden02}  A condition $\omega_\mathrm{ph}<\Delta$ for the conventional spin-Peierls mechanism is fulfilled in the organic compounds, but not in CuGeO$_3$.  Therefore, the spin-Peierls transition in CuGeO$_3$ can be explained by the theory in the antiadiabatic limit.
 
CuGeO$_3$ has an antiferromagnetic NN interaction $J_1$ and an antiferromagnetic NNN interaction $J_2$ that induces frustration.  Experimental and theoretical studies have shown that $J_1\approx 100 \,\mathrm{K}$ and $\alpha\cong0.36$.~\cite{nishi94,castilla95,riera95}  The instability of the spin-Peierls phase due to $J_2$ has been discussed by Wei{\ss}e~\cite{weisse99}.

Spin excitations in CuGeO$_3$ have been observed by inelastic neutron scattering.~\cite{arai96,braden99,nakamura09}  The separation of spin and phonon excitations in the experiments has not been complete, although some of phonon excitations have been identified by the same technique.~\cite{braden98-1,braden98-2,braden02} A recent development of polarized neutron scattering may resolve this problem in the near future. To give theoretical supports for inelastic polarized neutron scattering experiment, we investigate spin excitation of the spin-Peierls model with the non-softening phonon.  Our chief aim in the present paper is to clarify the effect of non-softening phonon on spin excitations in CuGeO$_3$.

We use the dynamical density-matrix renormalization-group (D-DMRG) method to calculate dynamical spin correlation function and phonon excitation spectrum.  This method is a dynamical version of DMRG method~\cite{white93} presented by Jeckelmann~\cite{jeckelmann02}. 

In this paper, we perform D-DMRG calculation of the dynamical spin correlation function for both the spin-Peierls model and its effective spin model ($J_1$-$J_2$ model). Treating phonons in the spin-Peierls model as quantum objects, we find a new spin excitation assisted by non-softening phonon. There is no corresponding structure in the effective $J_1$-$J_2$ model. The new structure shows a dispersive feature with strong intensity near the momentum $\pi$, and it is located above the phonon energy.  The new structure is explained by using particle-hole excitation assisted by phonon in the Su-Schrieffer-Heeger model onto which the spin-Peierls model with $XY$-type spin is mapped. The phonon excitation spectrum is also influenced by the spin-phonon interaction. Main peaks of phonon near the momentum $\pi$ show a new tail structure whose energy range is the same as the new spin excitation. In addition, spin-assisted phonon structures appear at low-energy region. These features induced by spin-phonon interaction are expected to be observed in inelastic neutron scattering experiments in the near future. 

This paper is organized as follows. In Sec.~\ref{Model}, we show a model Hamiltonian of the frustrated spin-Peierls chain with Einstein phonon and introduce a renormalized $J_1$-$J_2$ Hamiltonian.  We explain the numerical method, D-DMRG, in Sec.~\ref{Method}.  In preparation for the spin-Peierls model, we demonstrate dynamical spin correlation function in the $J_1$-$J_2$ model in Sec.~\ref{J1J2}. Dynamical spin correlation function and phonon excitation spectrum of the spin-Peierls model are presented in \S\ref{SP}.  Summary of the present paper is given in Sec.~\ref{Summary}.

\section{Model}
\label{Model}
We consider the following model Hamiltonian that describes 1D frustrated spin-Peierls chain.
\begin{equation}
\mathcal{H}=\mathcal{H}_\mathrm{s}+\mathcal{H}_\mathrm{p}+\mathcal{H}_\mathrm{sp}
\label{TotalH}
\end{equation}
with
\begin{eqnarray}
\mathcal{H}_\mathrm{s}&=&J\sum_j h_j^{(1)}(\Delta)+\alpha_0J\sum_j h_j^{(2)}(\Delta), \label{Hs}\\
\mathcal{H}_\mathrm{p}&=&\sum_j\frac{p_j^2}{2M}+\frac{K}{2} x_j^2, \label{Hp}\\
\mathcal{H}_\mathrm{sp}&=&g\sum_j\left( x_j-x_{j+1} \right)h_j^{(1)}(\Delta), \label{Hsp}
\end{eqnarray}
and
\begin{equation}
h_j^{(r)}(\Delta)=\bm{S}_j\cdot\bm{S}_{j+r}-\Delta S_j^zS_{j+r}^z.
\end{equation}
where $\bm{S}_j$ is the $j$-site spin operator for $S=\frac{1}{2}$, $x_j$ is the $j$-site displacement of the coordinate with respect to equilibrium position, and $p_j$ is the conjugate momentum operator.  $J$ is bare NN exchange interaction, and $\alpha_0$ is the ratio of bare NN and NNN interactions.  In this work, we consider only antiferromagnetic coupling for the NN and NNN interactions.
$M$ and $K$ are the effective mass and the elastic coupling constant, respectively. $g$ is the spin-phonon coupling constant.  $\Delta$ is the Ising anisotropic parameter.

After the second quantization of lattice degree of freedom, the phonon part of the Hamiltonian (\ref{Hp}) and the spin-phonon coupling term (\ref{Hsp}) are rewritten by
\begin{equation}
\mathcal{H}_\mathrm{p}= \omega_0\sum_jb_j^\dagger b_j
\end{equation}
and
\begin{equation}
\mathcal{H}_\mathrm{sp}=\frac{\lambda J}{2}\sum_j\left( b_j+b_j^\dagger-b_{j+1}-b_{j+1}^\dagger \right)h_j^{(1)}(\Delta) \label{eq:hsp2},
\end{equation}
respectively, where $b_j^\dagger$ and $b_j$ are the creation and annihilation operator of $j$-site phonon, respectively.  We employ the Einstein phonon with frequency $\omega_0=\sqrt{K/M}$.  The dimensionless spin-phonon coupling constant $\lambda$ is defined as $\lambda=(g/J)\sqrt{2/M\omega_0}$. We use the natural unit, $\hbar=1$.

The spin-Peierls model without the NNN interaction in the $XY$ limit ($\alpha_0=0$ and $\Delta=1$), that we call $XY$ spin-Peierls model, can be mapped to the Su-Schrieffer-Heeger (SSH) model~\cite{su79} by the Jordan-Wigner transformation,
\begin{equation}
\mathcal{H}_\mathrm{e}=-J\sum_k \cos(k) \tilde{c}_k^\dagger \tilde{c}_k \label{Hs_ssh}
\end{equation}
and
\begin{equation}
\mathcal{H}_\mathrm{ep}=i\lambda J\sum_{k,l} \left[\sin(k)-\sin(l)\right] (\tilde{b}_{k-l}+\tilde{b}_{l-k}^\dagger) \tilde{c}_k^\dagger \tilde{c}_l, \label{Hsp_ssh}
\end{equation}
where $\tilde{b}_q$ and $\tilde{b}_{q}^\dagger$ are the momentum representation of the phonon creation and annihilation operators. $\tilde{c}_k^\dagger$ and $\tilde{c}_k$ are the momentum representation of the spinless-charge creation and annihilation operators.

An effective Hamiltonian with Heisenberg spin exchange interactions ($\Delta=0$) after renormalization of phonon degree of freedom by the flow equation is given by~\cite{uhrig98}
\begin{equation}
\mathcal{H}_{J_1J_2}=J_1\sum_{j=1}^N \mathbf{S}_j\cdot\mathbf{S}_{j+1}+J_2\sum_{j=1}^N \mathbf{S}_j\cdot\mathbf{S}_{j+2} \label{eq:hamil_j1j2}
\end{equation}
with
\begin{eqnarray}
J_1&=&J\left[1+\frac{\lambda^2J}{4\omega_0}-\frac{3(1-\alpha_0)\lambda^2J^2}{8\omega_0^2}\right]+O(\lambda^3),\label{eq:fe_j1}\\
J_2&=&J\left[\alpha_0+\frac{\lambda^2J}{8\omega_0}+\frac{(3-5\alpha_0)\lambda^2J^2}{8\omega_0^2}\right]+O(\lambda^3),\label{eq:fe_j2}    
\end{eqnarray}
where $\lambda J/\omega_0$ and $J/\omega_0$ are small parameters for expansion and $O(\lambda^3)$ is the third or higher order terms of $\lambda$.  Equation~(\ref{eq:hamil_j1j2}) is a $J_1$-$J_2$ model with effective NN and NNN interactions. In this model, with increasing the ratio $\alpha=J_2/J_1$ from zero, the KT transition from spin liquid phase to dimer phase occurs at $\alpha_c$.~\cite{affleck89,eggert96,okamoto92}  In the spin-liquid phase where $\alpha<\alpha_c$, a gapless excitation appears.  On the other hand, in the dimer phase where $\alpha>\alpha_c$, there is a spin gap between the ground state and the lowest triplet excitation, accompanied by spontaneously dimerization of spin pairs.  The quantum phase transition does not require any soft phonon.

\section{Method}
\label{Method}
To examine dynamical behavior for the spin-Peierls model, we calculate two quantities, dynamical spin correlation function and phonon excitation spectrum.  

The dynamical spin correlation function at zero temperature is given by
\begin{equation}
\chi_s(k,\omega)=\Im \frac{1}{\pi L}\langle 0|S^z(k)\frac{1}{\omega+\epsilon_0-\mathcal{H}+i\gamma}S^z(k)|0\rangle, \label{dsf}
\end{equation}
where $L$ is the system size, $\gamma$ is the damping factor with small positive number, and $|0\rangle$ is the ground state.  $S^z(k)$ is the $z$ component of the spin operator at momentum $k$. 

We define phonon excitation spectrum as
\begin{equation}
\chi_p(q,\omega)=-\Im \frac{1}{\pi L}\langle 0|b(q)\frac{1}{\omega+\epsilon_0-\mathcal{H}+i\gamma}b^\dagger(q)|0\rangle,\label{pes}
\end{equation}
where $b(q)$ and $b^\dagger(q)$ are the phonon annihilation and creation operators at momentum $q$, respectively.  In eq.~(\ref{pes}), we consider a pair of $b(q)$ and $b^\dagger(q)$. In the phonon Green's function, there are additionally three pairs of $b^\dagger(q)$ and $b(q)$, $b(q)$ and $b(q)$, and $b^\dagger(q)$ and $b^\dagger(q)$. If the spin-phonon coupling is small, there are few phonons in the ground state. In such a case, the contribution from the three pairs is expected to be small. Therefore, we consider only the case of eq.~(\ref{pes}) to represent the phonon excitation spectrum.

For strongly-correlated 1D systems, it is well-known that the DMRG method can provide a good numerical solution of the ground state.~\cite{white93}  The dynamically extended version of DMRG, D-DMRG, is also suitable for obtaining dynamical properties at zero temperature.~\cite{jeckelmann02} Since the spin-phonon interaction (\ref{eq:hsp2}) breaks the mirror symmetry, we consider two reduced density matrices for both the system and the environment blocks in the DMRG process. We use the infinite-size algorithm of DMRG~\cite{white93} for systems with open boundary condition (OBC). In OBC, the Fourier transform of the spin and phonon operators read
\begin{eqnarray}
S^z(k)=\sqrt{\frac{2}{L+1}}\sum_{j=1}^{L}S_j^z\sin(jk),\label{Sz}\\
b(q)=\sqrt{\frac{2}{L+1}}\sum_{j=1}^{L}b_j^z\sin(jq),
\end{eqnarray}
with momenta $k$ and $q$ given by $n\pi/(L+1)$, $(n=1,2,\cdots,L)$.

For the D-DMRG method, we use three target states: $|0\rangle$, $S^z(k)|0\rangle$, and $\left[\omega+\epsilon_0-\mathcal{H}+i\gamma\right]^{-1}S^z(k)|0\rangle$ for eq.~(\ref{dsf}), and $|0\rangle$, $b^\dagger(q)|0\rangle$, and $\left[\omega+\epsilon_0-\mathcal{H}+i\gamma\right]^{-1}b^\dagger(q)|0\rangle$ for eq.~(\ref{pes}).  In the D-DMRG procedure, we use a modified version of the conjugate gradient method to calculate the correction vector, $\left[\omega+\epsilon_0-\mathcal{H}+i\gamma\right]^{-1}\hat{\mathcal{O}}|0\rangle$.  

\section{Results}
\label{Results}
In this section, we first show calculated results of dynamical spin correlation function for the effective $J_1$-$J_2$ model (\ref{eq:hamil_j1j2}) deduced from the spin-Peierls Hamiltonian (\ref{TotalH}). Secondly, treating phonons quantum-mechanically, we calculate the dynamical spin correlation function of the spin-Peierls model (\ref{TotalH}). New structures originated from the quantum phonons are identified by making a comparison with the $J_1$-$J_2$ model. Finally the effect of spin-phonon coupling on phonon excitation spectrum is examined.

In order to simulate CuGeO$_3$, we take the phonon energy $\omega_0=3J$,~\cite{nishi94,braden98-1,braden98-2,braden02} except in the case explicitly provided. The value of spin-phonon coupling is not clear for CuGeO$_3$. Therefore, we take $\lambda$ satisfying $\lambda J/\omega_0 <1$.

\subsection{Effective $J_1$-$J_2$ model}
\label{J1J2}
\begin{figure}
\centering
\includegraphics[scale=0.7]{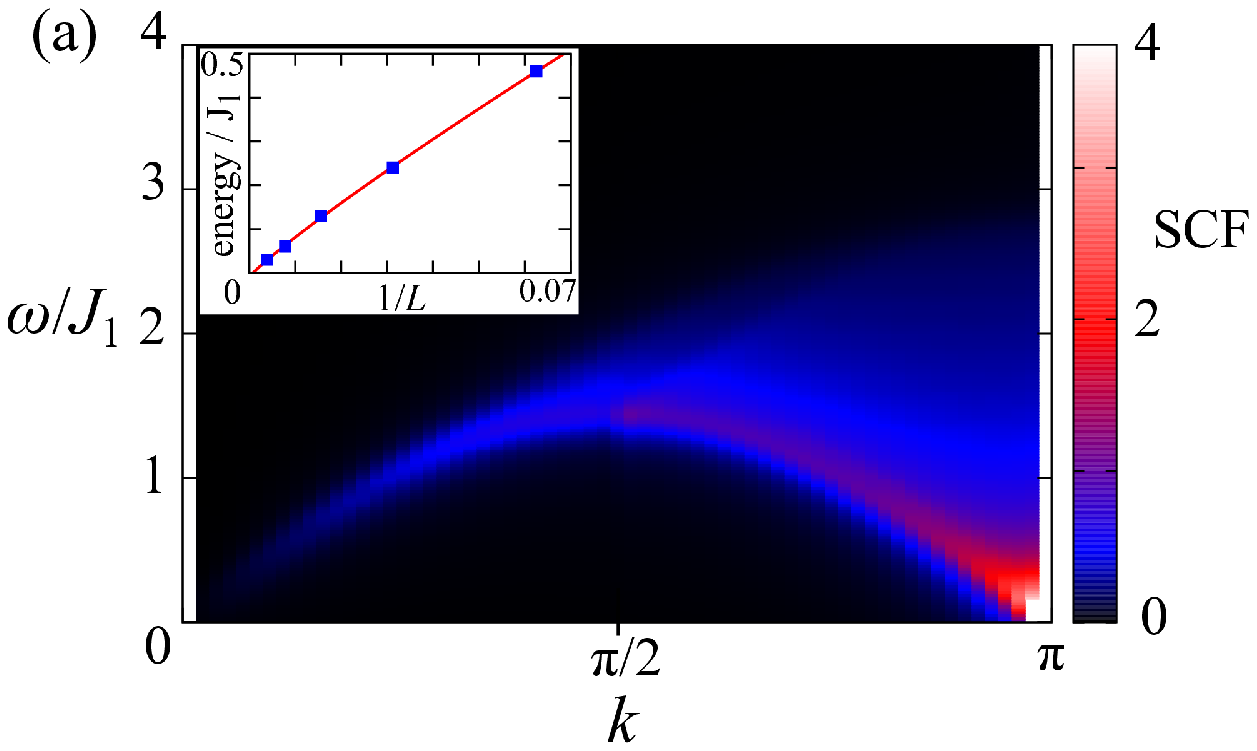}
\includegraphics[scale=0.7]{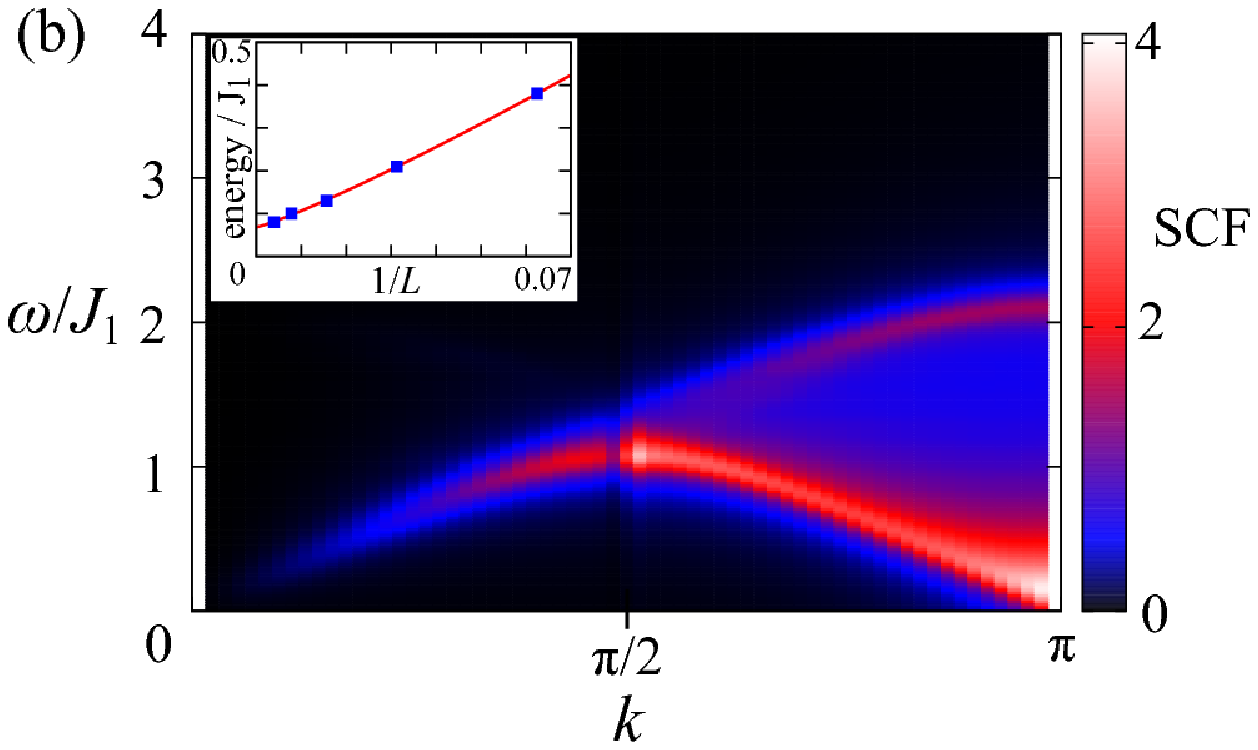}
\caption{\label{fig:dsf_zgzg} (Color online) Intensity map of dynamical spin correlation function (SCF) in a 64-site $J_1$-$J_2$ chain under two conditions: (a) $J_2/J_1=0.1$ ($\omega_0/J=3, \alpha_0=0, \lambda\cong1.12$), (b) $J_2/J_1=0.4$ ($\omega_0/J=3, \alpha_0=0.36, \lambda\cong1.06$).  The inset in both panels shows system-size dependence of the energy of peak position at the smallest momentum $k=\pi/(L+1)$.  Blue squares represent calculated peak positions, and red lines denote fitting by a function with a power: $f(1/L)=a(1/L)^b+c$.  $a=5.63$, $b=0.90$, and $c=-0.01$ for (a), and $a=7.43$, $b=1.14$, and $c=0.07$ for (b). The gap in (b) remains finite in the thermodynamical limit $1/L\rightarrow 0$.  The truncation number of DMRG and the broadening factor of Lorentzian are set to be $m=100$ and $\gamma=0.1J$, respectively.}
\end{figure}

The dynamical spin correlation function for a 64-site $J_1$-$J_2$ Heisenberg chain is shown in Fig.~\ref{fig:dsf_zgzg}. The damping factor $\gamma$ in eq.~(\ref{dsf}) is taken to be $0.1J$. We note that a preliminary result has been reported in ref.~32). 
In Fig.~\ref{fig:dsf_zgzg}(a), we take $\alpha_0=0$ and $\lambda=1.12$, resulting in $J_2/J_1=0.1$ from eqs.~(\ref{eq:fe_j1}) and (\ref{eq:fe_j2}). The ratio is below $\alpha_c$. This means that Fig.~\ref{fig:dsf_zgzg}(a) represents the dynamical spin correlation in a spin-liquid phase. The distribution of spectral weight is similar to the exact results of the 1D Heisenberg model ($J_2/J_1=0$) where spectral weight consists of the des Cloizeaux-Pearson mode at the lowest-energy branch and multi-spinon continuum.~\cite{bougourzi96,karbach97,biegel02} The inset in Fig.~\ref{fig:dsf_zgzg}(a) shows the system-size dependence of the position of the des Cloizeaux-Pearson mode at the smallest momentum $k=\pi/(L+1)$. A fitting function gives nearly zero excitation energy at $1/L\rightarrow 0$ as expected in the spin-liquid phase.

The spin correlation function in a dimer phase is shown in Fig.~\ref{fig:dsf_zgzg}(b), where $\alpha_0=0.36$,~\cite{nishi94,castilla95,riera95} $\lambda=1.06$, and thus $J_2/J_1=0.4$. We can find two characteristics as compared with Fig.~\ref{fig:dsf_zgzg}(a), i.e., strong intensity around $k=\pi/2$ ($\omega/J\sim 1$) and a peak structure at the upper edge of the spinon continuum around $k=\pi$ ($\omega/J\sim 2$). These features in the dimer phase have been reported in a previous study for small systems up to 16 sites under periodic boundary condition.~\cite{yokoyama97,muller81} The inset shows the presence of spin gap in the thermodynamic limit. The gap magnitude ($\sim 0.08J$) is similar to a previous report evaluated from DMRG calculations.~\cite{white96} These results also confirm the validity of our D-DMRG calculation.

There is a breakpoint near $k=\pi/2$ in the lowest-energy branch in Fig.~\ref{fig:dsf_zgzg}(b). This is an artifact by finite-size effect under OBC, since momenta near $k=\pi/2$ have a substantial contribution from the edges of the system as expected from eq.~(\ref{Sz}). This, however, does not occur for momenta close to $k=0$ and close to $k=\pi$, for which $\sin(jk)$ in eq.~(\ref{Sz}) goes to zero with approaching the edges of the system.

\subsection{Spin-Peierls model}
\label{SP}
We calculate dynamical spin correlation function and phonon excitation spectrum for a 16-site frustrated spin-Peierls chain in this section. The system size is smaller than that for the $J_1$-$J_2$ model (64 sites).  This is because single-site dimension of spin-phonon coupled system is several times larger than that of pure spin system, resulting in the requirement of huge computational resources. 

We checked the convergence of the calculation in terms of two parameters, i.e., the DMRG truncation number, $m$, and the maximal phonon number per site, $n_p$. We found that good convergence not only for the ground state but also for spectral weight at high-energy region can be achieved for $m=100$ and $n_p=2$.  In this paper, we use the broadening factor for the dynamical quantities $\gamma=0.1J$.

\begin{figure}
\centering
\includegraphics[scale=0.6]{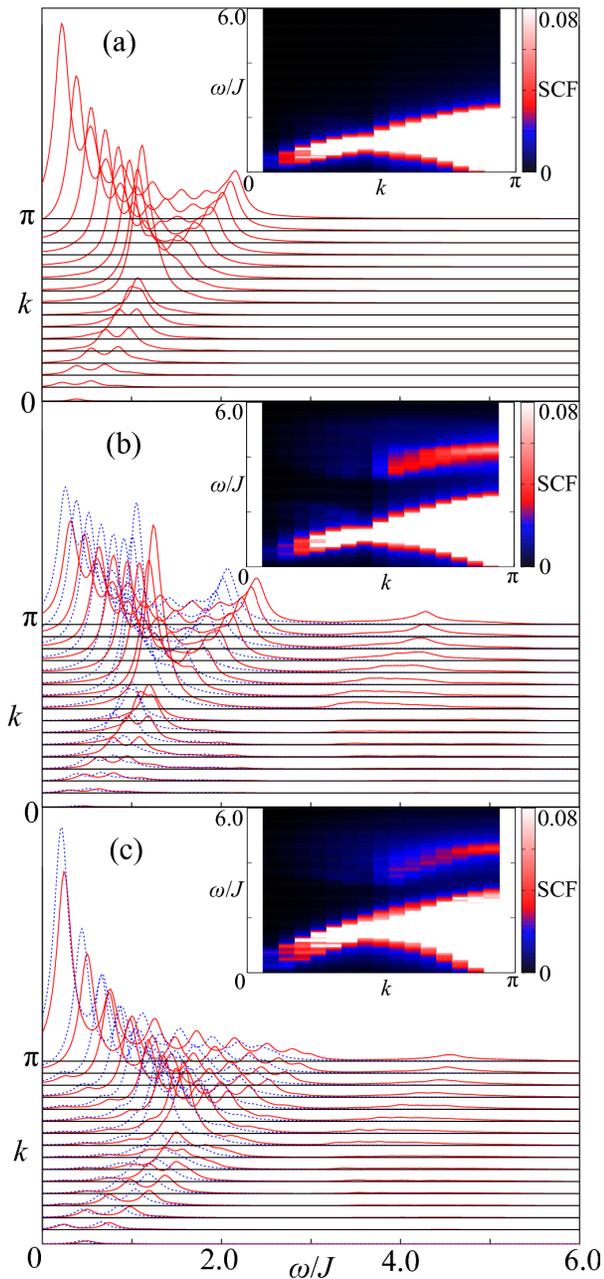}
\caption{\label{fig:dsf_sp_w3.0} (Color online) Dynamical spin correlation function (SCF) in a 16-site spin-Peierls chain. The phonon energy $\omega_0=3J$. (a) $\alpha_0=0.36$, $\lambda J/\omega_0=0$, (b) $\alpha_0=0.36$, $\lambda J/\omega_0=0.5$, and (c) $\alpha_0=0$, $\lambda J/\omega_0=0.5$. The blue dotted lines in (b) and (c) show the dynamical spin correlation function of the effective $J_1$-$J_2$ model, where $J_1$ and $J_2$ are evaluated from eqs.~(\ref{eq:fe_j1}) and (\ref{eq:fe_j2}), respectively: $J_2/J_1=0.44$ in (b) and $J_2/J_1=0.17$ in (c).  The inset in each panel shows the intensity map of the correlation function. The truncation number of DMRG is set on $m=100$ and maximal phonon number is two. The broadening factor is set to be $\gamma=0.1J$.}
\end{figure}

\subsubsection{Dynamical spin correlation function}
\label{DSCF}
Figure~\ref{fig:dsf_sp_w3.0}(a) shows the case without spin-phonon coupling, i.e., only $\mathcal{H}_s$ term in eq.~(\ref{Hs}) with $\alpha_0=0.36$. This is nothing but a 16-site result of the $J_1$-$J_2$ model, and the spectral behavior is consistent with the 64-site result with spin gap mentioned in \S\ref{J1J2}. 

The introduction of $\lambda$ changes the spectrum as shown in Fig.~\ref{fig:dsf_sp_w3.0}(b), where $\lambda J/\omega_0=0.5$. The most striking change is the appearance of high-energy spectral weight around $\omega\sim 4J$ (see red solid lines and intensity map shown in the inset). The intensity is strongest at $k=\pi$ and widens toward low-energy side with decreasing $k$ followed by an energy minimum at $k=\pi/2$. The energy position is higher than the dispersionless phonon located at $\omega_0=3J$. From this result, it is expected that additional spin excitations originating from the spin-phonon coupling exist above the energy of the phonons. Note that the intensity of the new structure increases with increasing $\lambda$. 

Comparing red solid lines in Fig.~\ref{fig:dsf_sp_w3.0}(b) with those in Fig.~\ref{fig:dsf_sp_w3.0}(a), we find a change of spectral distribution along the lowest energy branch: the spectral intensity at $k=\pi$ is suppressed and the weight transfers toward $k=\pi/2$. This change is similar to that caused by increasing $J_2/J_1$. This is reasonable, since the effective value of $J_2/J_1$ evaluated from the second-order expression of $\lambda$ in eqs.~(\ref{eq:fe_j1}) and (\ref{eq:fe_j2}) is $\alpha=J_2/J_1=0.44$, which is larger than the bare value $\alpha_0=0.36$. For comparison, the spin correlation function for the $J_1$-$J_2$ model with $\alpha=0.44$ is plotted in Fig.~\ref{fig:dsf_sp_w3.0}(b) as blue dotted lines. The lowest-energy branch follows that of the spin-Peierls model. This tempts us to justify the use of effective $J_1$-$J_2$ model. However, we can find a qualitative difference between the $J_1$-$J_2$ model and the spin-Peierls model: the upper edge of multi-spinon excitation ($\sim2.5J$ near $k=\pi$) increases in the spin-Peierls model while decreases in the $J_1$-$J_2$ model as compared with that of Fig.~\ref{fig:dsf_sp_w3.0}(a). A simple explanation of the difference would be that the second-order contributions in eqs.~(\ref{eq:fe_j1}) and (\ref{eq:fe_j2}) are not enough for the complete description of the spin-Peierls model and higher-order terms contributes significantly for the present parameter set. 

Figure~\ref{fig:dsf_sp_w3.0}(c) shows the case of $\alpha_0=0$ but with the same $\lambda$ as Fig.~\ref{fig:dsf_sp_w3.0}(b). As is the case of Fig.~\ref{fig:dsf_sp_w3.0}(b), there appears a high-energy dispersive structure with small intensity around $\omega=4J$ induced by the spin-phonon coupling in the spin-Peierls model. The upper edge of multi-spinon excitations ($\omega<3J$) in the spin-Peierls model is larger than that of the effective $J_1$-$J_2$ model. This is again the same as Fig.~\ref{fig:dsf_sp_w3.0}(b), indicating insufficient mapping of the spin-Peierls model onto the $J_1$-$J_2$ model for the present parameter set. We find small spectral weights below the des Cloizeaux-Pearson mode. The weights are caused by finite-size effect and decrease with increasing system size.

\begin{figure}
\centering
\includegraphics[scale=0.65]{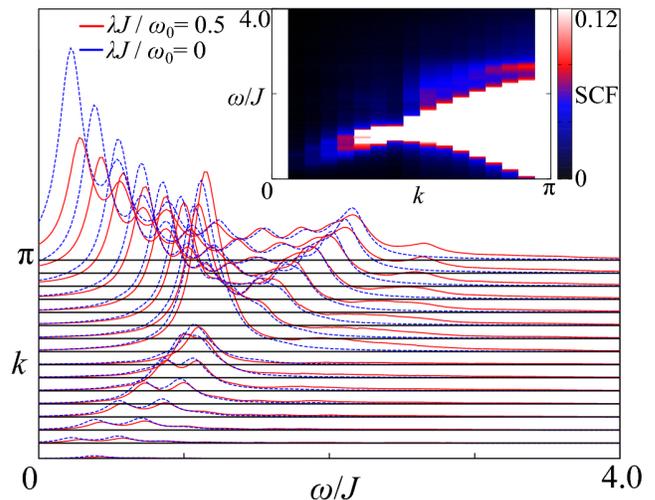}
\caption{\label{fig:dsf_sp_w1.5} (Color online) Dynamical spin correlation function in a 16-site spin-Peierls chain with $\alpha_0=0.36$ and $\omega_0/J=1.5$. The red lines represent the case of $\lambda J/\omega_0=0.5$, while the blue dotted lines represent the case without the coupling. The inset shows intensity map of the correlation function for $\lambda J/\omega_0=0.5$.}
\end{figure}

In order to examine the case where phonons are inside the multi-spinon continuum, we take $\omega_0/J=1.5$ (refs.~23-25) 
and show the spin correlation function for $\lambda J/\omega_0=0.5$ (red solid lines) in Fig.~\ref{fig:dsf_sp_w1.5}. Making a comparison with the case without the coupling (blue dotted lines), we find that phonon-induced spin excitations appear at $\omega\sim 2.5J$ with a dispersive structure showing a minimum at $k=\pi/2$. We note that, although the coupling constant $\lambda=0.75$ is smaller than the cases of Fig.~\ref{fig:dsf_sp_w3.0}(b) ($\lambda=1.5$), the spectral intensity relative to multi-spinon continuum is comparable to Fig.~\ref{fig:dsf_sp_w3.0}(b). This probably comes from the enhancement of hybridization between spin and phonon due to overlapping of their energy scale.

\begin{figure}
\centering
\includegraphics[scale=0.65]{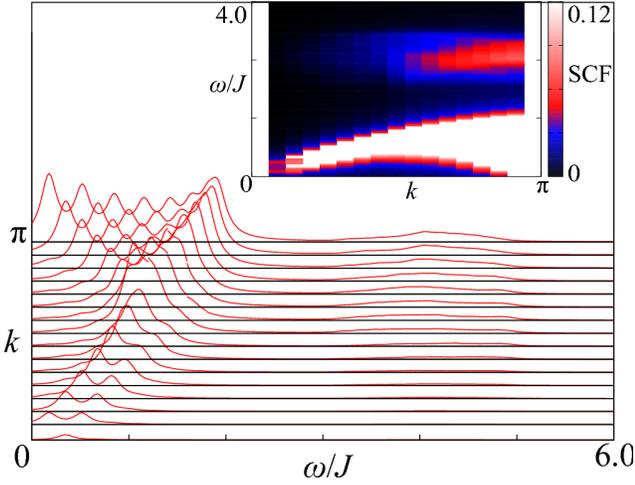}
\caption{\label{fig:dsf_XY_w3} (Color online) Dynamical spin correlation function in a 16-site $XY$ spin-Peierls chain (or dynamical charge-charge function in a 16-site Su-Schrieffer-Heeger chain).  The inset shows intensity map of the correlation function for $\lambda J/\omega_0=0.5$ and $\omega_0/J=3.0$.}
\end{figure}

Since the phonon-induced spin excitation appears in broad energy range, it is expected to be coupled with the multi-spinon continuum.  However, in the dimer phase of the spin-Peierls models, there is not only the continuum but also the state with large weight at the lower edge of the continuum. In order to make clear how the continuum contributes to the phonon-induced spin excitation, we examine the $XY$ spin-Peierls model ($\alpha_0=0$ and $\Delta=1$ in eqs.~(\ref{Hs}) and ~(\ref{eq:hsp2})), where the $XY$ spin chain has only continuum excitations of spinons.~\cite{muller81} Figure~\ref{fig:dsf_XY_w3} shows the dynamical spin correlation function for the $XY$ spin-Peierls model with $\lambda J/\omega_0=0.5$ and $\omega_0/J=3.0$.  We can see the same structure induced by the spin-phonon coupling in the high energy region as the case of the spin-Peierls model. Therefore, we conclude that the phonon-induced spin excitation couples with the continuum of spinons. 

\begin{figure}
\centering
\includegraphics[scale=0.6]{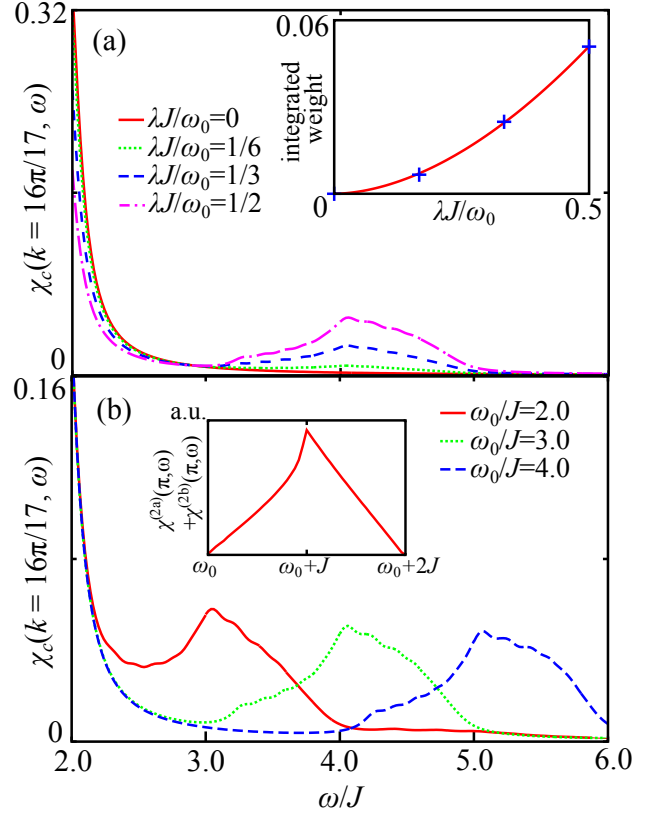}
\caption{\label{fig:dsf_XY_walp} (Color online) Dynamical spin correlation function in a 16-site $XY$ spin-Peierls chain at $k=16\pi/17\cong\pi$.  (a) the $\lambda$ dependence. The inset shows the $\lambda$ dependence of the integrated weight in the energy range of $3J$ and $6J$ for $\omega_0/J=3.0$. The blue crosses represent the integrated weight obtained by subtracting the integrated weight at $\lambda=0$, and the red line denotes the power-law fitting, $a(\lambda J/\omega_0)^b$ with $a=4.4\times 10^{-2}$ and $b=1.8$.  (b) $\omega_0$ dependence for $\lambda J/\omega_0=0.5$. The inset represents $\chi_c^{(2a)}(\pi,\omega)+\chi_c^{(2b)}(\pi,\omega)$ obtained by eqs.~(\ref{eq:chic2a}) and (\ref{eq:chic2b}).
}
\end{figure}

Moreover, we investigate the $\lambda$ and $\omega_0$ dependence of the phonon-induced spin excitation in the $XY$ spin-Peierls model. We show $\lambda$ dependence of the dynamical spin correlation function near $k=\pi$ with $\lambda J/\omega_0=0, 1/6, 1/3$, and $1/2$ for fixed $\omega_0/J=3.0$ in Fig.~\ref{fig:dsf_XY_walp} (a), and $\omega_0$ dependence with $\omega_0/J=2.0, 3.0$, and $4.0$ for fixed $\lambda J/\omega_0=0.5$ in Fig.~\ref{fig:dsf_XY_walp} (b).  We find that the energy position of the excitation does not depend on the spin-phonon coupling $\lambda$, but the integral of the excitation depends on $\lambda$ with a power-law behavior as seen in the inset of Fig.~\ref{fig:dsf_XY_walp}(a). In Fig.~\ref{fig:dsf_XY_walp}(b), we can see that the excitation is situated between $\omega_0$ and $\omega_0+2J$ at $k\cong\pi$.

The behaviors in the dynamical spin correlation function in the $XY$ spin-Peierls model can be explained by the SSH model in eqs.~(\ref{Hs_ssh}) and (\ref{Hsp_ssh}). We assume that phonon creation and annihilation operators never have any finite expectation values in the ground state.  This assumption is supported by the well-known fact that the ground state is not in the dimer phase with the bond alternation for small coupling constant $\lambda$.~\cite{fradkin83} The spin correlation function is rewritten by the spinless charge-charge correlation function.  We obtain the imaginary part of the spinless charge-charge correlation function, $\chi_c(k,\omega)$, within the second-order perturbation in terms of $\lambda$:
\begin{equation}
\chi_c(k,\omega)=\chi_c^{(0)}(k,\omega)+\chi_c^{(2a)}(k,\omega)+\chi_c^{(2b)}(k,\omega)+O(\lambda^4) \label{eq:chic}
\end{equation}
with 
\begin{eqnarray}
\chi_c^{(0)}(k,\omega)&=&\sum_{l}\delta(\omega-\epsilon_{l+k}+\epsilon_l)\theta(-\epsilon_l) \theta(\epsilon_{l+k}) \label{eq:chic0}\\
\chi_c^{(2a)}(k,\omega)&=&\left(\frac{\lambda J}{\omega_0}\right)^2 \sum_{l q} \left(\frac{\omega_0 B(l+k,l+k+q)}{\omega_{0}-\epsilon_{l+k}+\epsilon_{l+k+q}}\right)^2 \nonumber \\
&\times&\theta(-\epsilon_l)\theta(\epsilon_{l+k})\theta(\epsilon_{l+k+q}) \nonumber \\
&\times&\delta(\omega-\omega_{0}-\epsilon_{l+k+q}+\epsilon_l) \label{eq:chic2a} \\
\chi_c^{(2b)}(k,\omega)&=&\left(\frac{\lambda J}{\omega_0}\right)^2 \sum_{l q} \left(\frac{\omega_0 B(l,l+q)}{\omega_{0}-\epsilon_{l}+\epsilon_{l+q}}\right)^2 \nonumber \\
&\times&\theta(-\epsilon_l)\theta(\epsilon_{l+k})\theta(-\epsilon_{l+q}) \nonumber \\
&\times&\delta(\omega-\omega_{0}-\epsilon_{l+k}+\epsilon_{l+q}), \label{eq:chic2b}
\end{eqnarray}
where $\epsilon_k=-J\cos(k)$ and $B(k,l)=\sin(k)-\sin(l)$. $\theta(x)$ denotes the step function, and we assume $\omega_0>2J$.   In this approximation, phonon-assisted particle-hole excitation starts from the second order of $\lambda$ and is given by $\chi_c^{(2a)}(k,\omega)$ and $\chi_c^{(2b)}(k,\omega)$.  Thus this excitation is expected to increase as the power law $\lambda^b$ with $b=2$.  This behavior explains the power-law behavior of the integrated intensity of the excitation with $(\lambda J/\omega_0)^{1.8}$ as shown in the inset of Fig.~\ref{fig:dsf_XY_walp}(a). 

$\chi_c^{(2a)}$ and $\chi_c^{(2b)}$ include the particle-hole excitation accompanied by a phonon with the energy of $\omega_0$. The particle-hole excitation can scan the full-energy range of charge excitation with the width of $2J$. Therefore, the phonon-induced spin excitation in the $XY$ spin-Peierls model is expected to be located between $\omega_0$ and $\omega_0+2J$, which is seen in Fig.~\ref{fig:dsf_XY_w3} and Fig.~\ref{fig:dsf_XY_walp}(b).  At $k=\pi$,by using eqs.~(\ref{eq:chic2a}) and (\ref{eq:chic2b}), we obtain a cusp-like structure as shown in the iset of Fig.~\ref{fig:dsf_XY_walp}(b), which is consistent with the results shown in in Fig.~\ref{fig:dsf_XY_walp}. We note that this cusp-like structure is also similar to the phonon-assisted magnon absorption observed in the 1D Mott insulator Sr$_2$CuO$_3$.~\cite{suzuura96,lorenzana97}  Thus we conclude that the full-energy-range scanning with particle-hole excitation accompanied by a phonon is important to understand the phonon-induced spin excitation.

\subsubsection{Phonon excitation spectrum}
\begin{figure}
\centering
\includegraphics[scale=0.6]{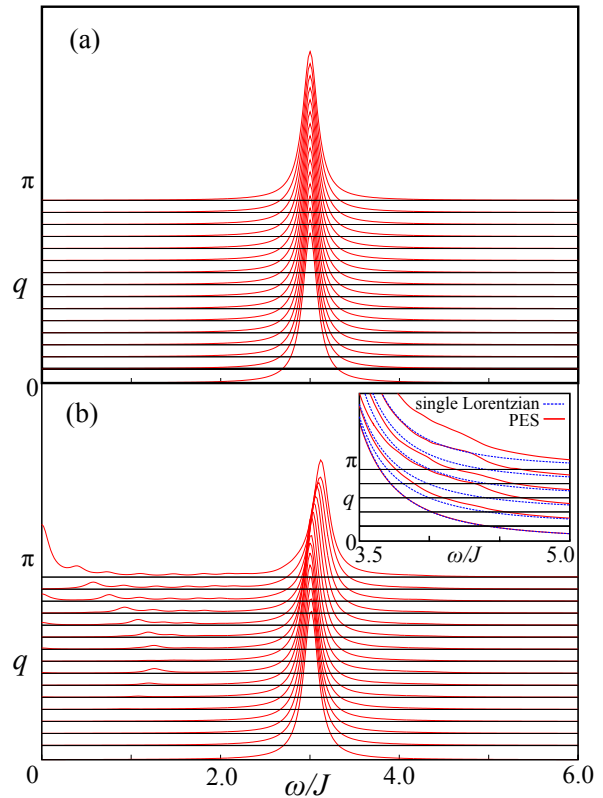}
\caption{\label{fig:dcf} (Color online) Phonon excitation spectrum in a 16-site spin-Peierls chain ($\alpha_0=0.36$, $\omega_0/J=3.0$) without spin-phonon coupling $\lambda J/\omega_0=0$ (a) and with the coupling $\lambda J/\omega_0=0.5$ (b). The inset shows the tail of main peak (red solid line) together with a single Lorentzian curve obtained by fitting the main peak (blue dotted line). The difference between the two lines in the inset demonstrates the presence of hump structure at the high-energy side of the main peak.}
\end{figure}

We examine how phonon excitations are influenced by spin-phonon coupling. Figure~\ref{fig:dcf}(a) shows the phonon excitation spectrum for the Einstein phonon with $\omega_0=3J$. The introduction of spin-phonon coupling with $\lambda J/\omega_0=0.5$ into a frustrated spin model ($\alpha_0=0.36$) changes the phonon excitation spectrum from Fig.~\ref{fig:dcf}(a) to Fig.~\ref{fig:dcf}(b). Two changes are seen in the phonon excitation spectra. One is a slight shift of the phonon main peak toward higher energy near $q=\pi$, accompanied by small hump at the high energy side as shown in the inset. The energy position of the hump structure is the same as that of phonon-induced spin excitations seen in Fig.~\ref{fig:dsf_sp_w3.0}(b). The other change is the emergence of low-energy phonon components ($\omega<1.5J$). The strongest change in intensity appears near $\omega=0$ at $q\sim \pi$.  We note that, since the phonon excitation occurs without spin flipping, i.e., in the Hilbert space of zero total spin, the energy positions of the spin-induced phonon excitations are different from those of spin excitations shown in \S\ref{SP}.  Both the high- and low-energy changes predominantly occur near $q=\pi$, which is due to the nature of spin-phonon coupling as shown in Appendix. We find that these changes are insensitive to the presence of $\alpha_0$ and are general features of the spin-phonon coupling.

\section{Summary}
\label{Summary}
We have investigated spin excitation for a spin-Peierls chain with nearest-neighbor and next-nearest-neighbor Heisenberg spin exchange interactions, taking CuGeO$_3$ into consideration.  We consider a gapped and dispersionless (Einstein) phonon as the lattice degree of freedom.  We then apply a dynamical density matrix renormalization group method to calculate dynamical spin correlation function at zero temperature.

We have found a new spin excitation assisted by non-softening phonon at the energy region higher than the phonon energy. The new spin excitation shows a dispersive feature with strong intensity near $k=\pi$. There is no corresponding structure in the effective $J_1$-$J_2$ model derived from the spin-Peierls model as expected.  This demonstrates the importance of treating the phonons quantum-mechanically. 

We have also found the new spin excitation appears in the $XY$ spin-Peierls model.  This model is equivalent to the Su-Schrieffer-Heeger model. We have shown that the behaviors of the new excitation are explained by charge-charge correlation function in the SSH model.

The phonon excitation spectrum is also influenced by the spin-phonon interaction. We have found the shift of main phonon peak toward higher energy side near $q=\pi$, accompanied by a new tail structure whose energy range is the same as the new spin excitation. In addition, new spin-assisted phonon structures appear at low-energy region. The fact that the strong modification occurs near $q=\pi$ is understood by taking into account the momentum dependence of the spin-phonon coupling. 

The effect of spin-phonon coupling on inelastic neutron scattering has been discussed so far in literature.~\cite{lovesey} However, there is no work demonstrating the intensity distributions for the spin-Peierls model as far as we know. Therefore, we believe that the present results will be helpful for analyzing inelastic neutron scattering data in spin-phonon coupled systems.

In CuGeO$_3$, inelastic neutron scattering experiments have clearly revealed the presence of both lowest-energy branch of spin excitation with strong intensity and multi-spinon continuum.~\cite{ arai96,braden99,nakamura09} This is consistent with the results of the $J_1$-$J_2$ model. Experimental data has also shown high-energy structures above the continuum. If the structures are originated from spin degrees of freedom, they may be due to the coupling with phonon. It is desired to resolve the structure into phonon and spin components by either detailed analyses of momentum dependence of their intensity or new polarized inelastic neutron scattering experiments. 


\section*{Acknowledgment}
We thank J. Kokalj and P. Prelov\v{s}ek for fruitful discussions. We also thank R. Kajimoto, K. Ikeuchi, F. Mizuno, M. Fujita, and M. Arai for discussions on inelastic neutron scattering data of CuGeO$_3$. 
This work was supported by Nanoscience Program of Next Generation Supercomputing Project, the Grant-in-Aid for Scientific Research (Grants No. 19052003 and No. 22340097) from MEXT, the Grant-in-Aid for the Global COE Program ``The Next Generation of Physics, Spun from Universality and Emergence,'' and the Yukawa International Program for Quark-Hadron Sciences at YITP, Kyoto University. A part of numerical calculations was performed in the supercomputing facilities in ISSP and ITC, the University of Tokyo, YITP and ACCMS, Kyoto University.  The financial support of JPSJ and MHEST under the Slovenia-Japan Research Cooperative Program is also acknowledged.

\appendix
\section{Momentum dependence of spin-phonon coupling}
To examine the nature of the spin-phonon coupling term (\ref{Hsp}), we introduce Holstein-Primakoff bosons assuming two sublattices. The corresponding boson operators are represented by $A_m$ and $B_m$ with integer $m$. We consider the Fourier transformation of the boson operators: $\tilde{a}_k=\sqrt{\frac{1}{L}}\sum_{j=1}^{N}e^{ijk}a_j$ with $a_{2m-1}=A_m$, $a_{2m}=B_m$.  By neglecting fourth- or more higher-order terms, the spin-phonon term (\ref{Hsp}) is rewritten by

\begin{equation}
 \mathcal{H}_\mathrm{sp}=\sum_q \mathcal{H}_\mathrm{sp}^\prime(q)
\label{A1}
\end{equation}
with
\begin{eqnarray}
\mathcal{H}_\mathrm{sp}^\prime(q)\cong i\frac{\lambda J}{2\sqrt{L}}\sum_k(\tilde{b}_q+\tilde{b}_{-q}^\dagger) \ \ \ \ \ \ \ \ \ \ \ \ \ \ \ \ \ \ \ \ \ \ \ \ \ \ \nonumber\\
\times \left[\sin(q)\tilde{a}_k^\dagger \tilde{a}_{k-q}+\sin(k) (\tilde{a}_k^\dagger \tilde{a}_{-k+q}^\dagger -\tilde{a}_k \tilde{a}_{-k-q})\right],
\end{eqnarray}
where $\tilde{b}_q^\dagger$ and $\tilde{b}_q$ are the creation and annihilation operators of phonons in the momentum representation.

We can find that the $q=0$ phonon has no coupling with spin in eq.~(\ref{A1}).  Actually, the term including the $q=0$-phonon operators reads
\begin{eqnarray}
\mathcal{H}_\mathrm{sp}^\prime(q=0)&=&i\frac{\lambda J}{2\sqrt{L}}(\tilde{b}_0+\tilde{b}_0^\dagger)\sum_{k} \sin(k) (\tilde{a}_k^\dagger \tilde{a}_{-k}^\dagger -\tilde{a}_k \tilde{a}_{-k}) \nonumber \\
&=&i\frac{\lambda J}{2\sqrt{L}}(\tilde{b}_0+\tilde{b}_0^\dagger)\sum_{k} \sin(-k) (\tilde{a}_k^\dagger \tilde{a}_{-k}^\dagger -\tilde{a}_k \tilde{a}_{-k}) \nonumber \\
&=&-\mathcal{H}_\mathrm{sp}^\prime(q=0)=0.
\end{eqnarray}
However, the interaction of the $q=\pi$ phonon and spin remains finite:
\begin{eqnarray}
\mathcal{H}_\mathrm{sp}^\prime(q=\pi)&=&i\frac{\lambda J}{2\sqrt{L}}(\tilde{b}_\pi+\tilde{b}_\pi^\dagger)\sum_{k}\cos(k) \nonumber \\
&\times& (\tilde{a}_{k+\pi/2}^\dagger \tilde{a}_{-k+\pi/2}^\dagger -\tilde{a}_{k+\pi/2} \tilde{a}_{-k+\pi/2}) \nonumber \\
&\neq & 0.
\end{eqnarray}
This is the reason why phonons near $q=\pi$ are affected strongly by the spin-phonon coupling as compared with the region near $q=0$.

\end{document}